\documentclass[amsmath,amssymb,aps,showpacs,twocolumn,floats,superscriptaddress]{revtex4}

\usepackage{graphicx}
\usepackage{amssymb}
\usepackage{amsmath}
\usepackage{hyperref}
\usepackage{footnote}
\usepackage{color}
\usepackage{multirow}
\usepackage{enumerate}
\usepackage{soul}
\usepackage{bbold}
\usepackage[FIGTOPCAP]{subfigure}
\pdfoutput=1 
\def \hH{ \hat{\mathcal{H}}} 
 
\newcommand{\ev}[1]{\ensuremath{\left\langle #1 \right\rangle}}

\newcommand{\PD}{\phantom{\dag}}

\newcommand{\IN}{\textup{in}}
\newcommand{\OUT}{\textup{out}}

\newcommand{\EFF}{\textup{eff}}
\newcommand{\QND}{\textup{QND}}
\newcommand{\ADD}{\textup{add}}
\newcommand{\HOP}{\textup{hop}}

\graphicspath{{./plots/}}

\begin{document}
   
\title{Quantum-limited amplification without instability}

\author{A. Metelmann}
\email[ Corresponding author: ]{anja.metelmann@kit.edu}
\affiliation{Dahlem  Center  for  Complex  Quantum  Systems  and  Fachbereich  Physik, Freie  Universit\"at  Berlin,  14195  Berlin,  Germany}
\affiliation{Institute for Theory of Condensed Matter, Karlsruhe Institute of Technology, 76131 Karlsruhe, Germany}
\affiliation{Institute for Quantum Materials and Technology, Karlsruhe Institute of Technology, 76344 Eggenstein-Leopoldshafen, Germany}
\author{O. Lanes}
\affiliation{Department of Physics and Astronomy, University of Pittsburgh, Pittsburgh, PA 15260, USA}
\author{T.-C. Chien}
\affiliation{Department of Physics and Astronomy, University of Pittsburgh, Pittsburgh, PA 15260, USA}
\author{A. McDonald}
\affiliation{Pritzker School of Molecular Engineering, University of Chicago, 5640 S. Ellis Ave., Chicago, IL 60637, USA} 
\affiliation{Department of Physics, University of Chicago, 5720 S. Ellis Ave., Chicago, IL 60637, USA}
\author{I. Tsiamis}
\affiliation{Dahlem  Center  for  Complex  Quantum  Systems  and  Fachbereich  Physik, Freie  Universit\"at  Berlin,  14195  Berlin,  Germany}
\author{M. Hatridge}
\affiliation{Department of Physics and Astronomy, University of Pittsburgh, Pittsburgh, PA 15260, USA}
\author{A.~A. Clerk}
\affiliation{Pritzker School of Molecular Engineering, University of Chicago, 5640 S. Ellis Ave., Chicago, IL 60637, USA}  
\date{\today}
\begin{abstract}
Quantum parametric amplifiers typically generate gain by operating in proximity to a point of dynamical instability.  We consider an alternate general strategy where quantum-limited, large-gain amplification is achieved without {\it any} proximity to a dynamical instability.  Our basic mechanism (involving dynamics that conserves the number of squeezed photons) enables the design of a variety of one and two mode amplifiers that are not limited by any fundamental gain-bandwidth constraint.  We focus on a particular realization that allows us to realize an ideal single-mode squeezing operation {\it in transmission}, which has zero reflection.  We present both a thorough theoretical analysis of this system (including pump-depletion effects), and also discuss results of an experimental superconducting quantum circuit implementation.    
\end{abstract}
\maketitle
%

\section{Introduction}

Parametric amplifiers having quantum-limited noise properties play a crucial role in a variety of quantum information technologies.  In optical-domain systems, they are a crucial resource for preparing both discrete-variable and continuous-variable entangled states \cite{WeedbrookRMP2012}.  For superconducting microwave circuits, quantum parametric amplifiers harnessing Josephson nonlinearities serve as a workhorse for fast, high-fidelity qubit measurements.  Given the crucial role they play, there has been an enormous amount of activity (especially in the microwave-domain) looking at alternate amplifier designs that provide advantages in terms of bandwidth, noise and isolation (i.e.,~in-built nonreciprocity)   
(see e.g.~Refs.~\cite{Castellanos-Beltran2008,Malnou2018,Bergeal2010, Bergeal2010a, Abdo2011, Abdo2013, Chien2019,Macklin2015,Sivak2019, Abdo2013b,Sliwa2015, Lecocq2017, Kamal2017, Mercier2019, Lecocq2019,Metelmann2014,Roy2015,Metelmann2015,Ockeloen2016, Metelmann2017,Zhong2019,Simonson2022}).  Almost all these strategies ultimately use parametric processes that induce dynamical instability: in the absence of external dissipation or additional nonlinearities, the internal intra-cavity system dynamics would lead to unbounded exponential growth.  This instability is cut-off by dissipation (i.e.,~the coupling to input and output transmission lines), and the underlying instability physics is used to generate the desired amplification.  This basic paradigm is at the heart of standard degenerate and non-degenerate parametric amplifier designs (see \cite{RevModPhys,Roy2016} for pedagogical reviews), as well as more complex amplifier designs.

While undeniably powerful, the strategy of harnessing an instability has some inherent drawbacks.  As we discuss, it necessarily leads to a fundamental gain-bandwidth tradeoff: amplification is achieved by tuning pump parameters closer and closer to instability, which correspondingly increases the system's response time.  This is analogous to the phenomenon of critical slowing down that occurs when one approaches a second-order phase transition.  The net result is a system bandwidth that scales inversely with the square-root of the amplifier's power gain \cite{RevModPhys,Roy2016}.  

\begin{figure} 
  \centering\includegraphics[width=\columnwidth]{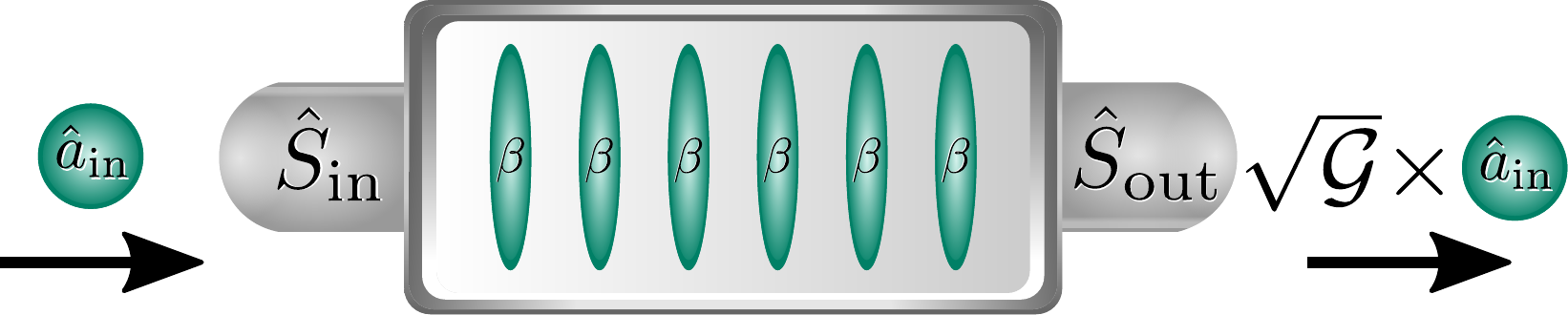} 
 	\caption{Photons entering the Bogoliubov amplifier at the input port are converted 
 	             into Bogoliubov quasiparticles via the squeezing transformation $\hat S_{\rm in}$.When leaving the amplifier they are converted back to photons via the inverse squeezing transformation $\hat S_{\rm out}$. Crucially, $\hat S_{\rm out} \neq \hat S_{\rm in}^{-1}$, results in the enhancement of the photon amplitude with factor $\sqrt{\mathcal G}>1$.} 
 	\label{Figure1}
\end{figure}

Here, we present an extremely general way to achieve ideal quantum limited amplification that does not involve proximity to a dynamical instability.  The basic idea is to exploit what at first glance seems a very sub-optimal situation:  use parametric processes that are effectively detuned from resonance.  Such systems are described by mean-field Hamiltonians that are fully stable even without any external dissipation.  They still however can generate remarkably ideal quantum-limited amplification: in particular, even though they use at most two photonic cavity modes, they are {\it fundamentally not subject to any gain-bandwidth constraints}.  As we discuss, this is a direct consequence of the eigenstates of these systems being {\it squeezed photons}.  Such squeezed photons are described by so-called Bogoliubov modes, bosonic annihilation operators that are generated by squeezing transformations acting on standard photonic creation and destruction operators.  The class of dynamically-stable amplifiers we introduce here all have the feature of having a set of Bogoliubov modes whose number operators are conserved.

The heuristic operation of our class of amplifiers is sketched in Fig. \ref{Figure1}.  Photons enter the input port, and are effectively converted to Bogoliubov quasiparticles; this involves a squeezing transformation $\hat {S}_{\rm in}$.  These quasiparticles then have simple, number conserving dynamics.  To leave the amplifier, these Bogoliubov quasiparticles are converted back to photons via a second squeezing transformation $\hat {S}_{\rm out}$.  Amplification is achieved by the simple fact that $ \hat {S}_{\rm out} \neq   \hat {S}_{\rm in}^{-1} $:  there is a net squeezing-amplification transformation implemented on photons scattered by the amplifier.   

Our schemes have the further advantage that by a simple parameter tuning, they can directly implement the enhanced bandwidth strategy of Ref.~\cite{Roy2015}, where the frequency dependent gain is extremely flat near resonance.  While Ref.~\cite{Roy2015} achieved this via the introduction of a secondary, optimized external impedance, in our systems, this bandwidth enhancement is in-built.  Yet another advantage of these designs is an intrinsic resilience against pump-depletion effects, something that is a direct consequence of not operating in proximity to an instability.  

We show that our strategy is extremely general, and discuss a variety of implementations.
The simplest corresponds to new way to operate a standard single-cavity degenerate parametric amplifier (DPA) such that there is no gain-bandwidth product;  this is analyzed in Sec.\ref{SecOptDetunedDPA}. 
We focus most of our attention however on an even more novel setup:  a two-mode, two-port amplifier that achieved a perfect DPA squeezing operation {\it in transmission}.  Despite the presence of an extra mode compared to a standard DPA, this system nonetheless has quantum-limited performance (i.e.,~one quadrature is amplified noiselessly).  It also has no gain-bandwidth limitation, and overcomes a key limitation of standard DPA: they operate in reflection, meaning that there is no intrinsic separation of amplifier input and output.  A detailed analysis of this setup is presented in Sec.\ref{SecOptImbalnecedDPA}. We also show that this setup is amenable with current superconducting circuit technology:  Sec.\ref{SecExpResults} describes results of an experimental implementation of this novel two-mode amplifier showing a bandwidth which is enhanced by over a factor of $6$ compared to the standard setups. Note that previous work has explored the use of detuned parametric driving to realize quantum non-demolition dynamics, which conserves one or more photonic quadratures \cite{Szorkovszky2013,Szorkovszky2014,Szorkovszky2014b}.  Such QND interactions are distinct from the ideas we present here; in particular such QND systems are on the cusp of instability (i.e.,~they cannot be diagonalized), where as our systems are fully stable (i.e.,~described by diagonalizable Hamiltonians).   


\section{The basics: single mode Bogoliubov-mode amplifier}


\subsection{Recap of a standard DPA} \label{SecRecapDPA}
We first recall the basics of a DPA in the stiff-pump limit. The amplifier consists of a principle cavity with a weak nonlinear coupling to an auxiliary pump mode. By driving this mode strongly with an external pump at an appropriate frequency, one realizes a mean-field Hamiltonian of the form ($\hbar = 1$): 
\begin{align}\label{eq:DPA}
    \hH
	=
	\Delta \hat a^\dagger \hat a
	+
	\frac{\nu}{2}
	\left(
	\hat a^\dagger \hat a^\dagger
	+
	\hat a \hat a
	\right) .
\end{align}
We are working in a rotating frame set by the pump frequency.  $\hat a$ is the photon lowering operator for the signal mode, and $\Delta$ is the detuning of the pump from the cavity resonance frequency. $\nu$ is the effective parametric drive amplitude (determined by both the nonlinearity and the pump amplitude).  We assume without loss of generality that both $\Delta$ and $\nu$ are real and positive.
Note that for $\Delta < \nu$, the DPA Hamiltonian given in Eq.~\ref{eq:DPA} is unstable:  it cannot be diagonalized, and in the absence of the dissipation it generates unbounded exponential growth, corresponding to a dynamical instability. 
    
Coupling the system to an input-output waveguide with coupling rate $\kappa$ makes the system dynamically stable as long as $\kappa/2 \geq\nu$. Gain is generated by approaching the point of instability, i.e. by increasing $\nu$ so it approaches $\kappa/2$ from below. Signals incident on the waveguide will be reflected with gain. 
   
The frequency dependent gain is obtained via input-output theory \cite{Gardiner1985} and yields for zero detuning $(\Delta = 0)$ 
\begin{align}\label{eq:Gain_Normal}
       \mathcal G[\omega] 
        =
        \frac{\left(\frac{2\omega}{D}\right)^2+ \mathcal G_0}{\left(\frac{2\omega}{D}\right)^2+1}, 
        \hspace{0.5cm}
         \sqrt{\mathcal G_0}   = \frac{\frac{\kappa}{2}+\nu}{\frac{\kappa}{2}- \nu} ,
\end{align}
where $D = 2 \kappa/( \sqrt{\mathcal G_0}+1)$ serves as the effective bandwidth of the amplifier, defined here as the full width at half of the maximum gain. Only signals contained within a frequency range of approximately $D$ around resonance  will be significantly amplified. In the relevant case where the zero frequency power gain is very large $ \mathcal G_0 \gg 1$, we have 
\begin{align}
        D \approx \frac{2\kappa}{ \sqrt{\mathcal G_0} } \implies D \sqrt{ \mathcal G_0} \approx 2 \kappa  .
\end{align}
This encapsulates the fundamental gain-bandwidth product that limits conventional parametric amplifiers. If one wants to increase the peak gain, it is necessarily accompanied by a reduction in the operating bandwidth of the amplifier.  This is a generic feature of any amplifier that generates gain by operating closer to a point of dynamical instability.   

\begin{figure} 
  \centering\includegraphics[width=0.45\textwidth]{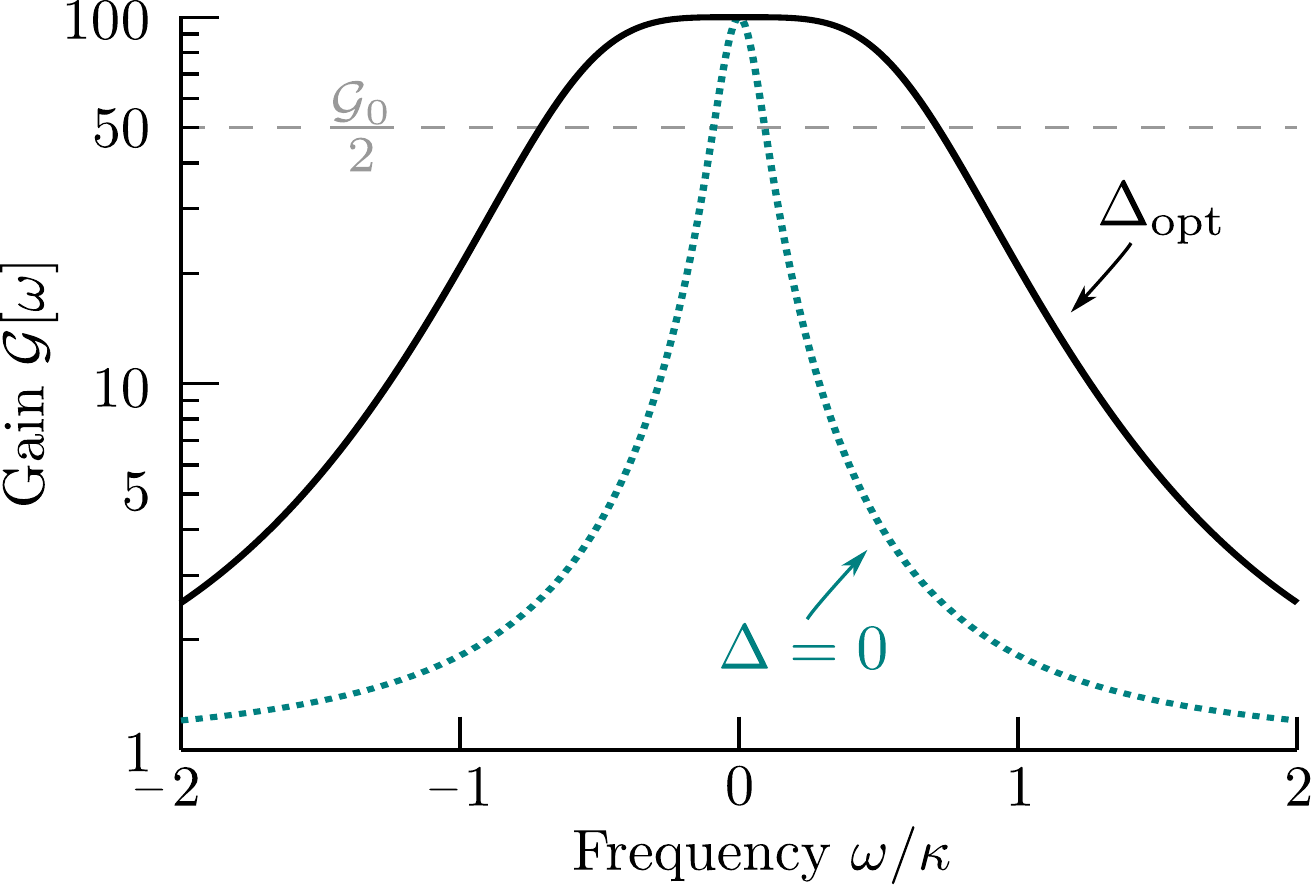}
 	\caption{
 	Plot of the gain as a function of injected signal frequency for a zero frequency gain of 20dB. The optimally detuned Bogoliubov amplifier (ODBA) has a broad range of frequencies over which amplification occurs (solid black line). Conventional parametric amplifiers only amplify in a narrow range of frequencies (dotted teak line).
 	}
 	\label{Figure2}
\end{figure}

\subsection{Optimally-detuned DPA: The ODBA}\label{SecOptDetunedDPA}
    
We now consider our DPA system in a case where the Hamiltonian is stable and diagonalizable.  This requires $\Delta \geq \nu$. Here, the Hamiltonian can be diagonalized as
    \begin{equation}
        \hH = \Lambda \; \hat {\beta}^\dagger \hat {\beta}, \,\,\,\,\,\,
        \hat {\beta} = \cosh(r) \hat {a} + \sinh(r) \hat {a}^\dag ,
    \end{equation}
with $\Lambda = \sqrt{\Delta^2 - \nu^2}$ and where $\hat {\beta}$ is a canonical bosonic lowering operator in the Bogoliubov basis. 
The eigenstates of the Bogoliubov mode number operator $\hat n_{r} = \hat\beta^{\dag}\hat \beta $ are squeezed Fock states $|n_{r}\rangle = \hat S(r) |n\rangle$ with $\hat S(r) = \exp{[r/2(\hat a \hat a - \hat a^{\dag} \hat{a}^\dag)]}$.
The squeezing parameter $r$ is given by $\tanh 2 r = \nu/\Delta $. The dynamics of this Hamiltonian in the Bogoliubov basis are obviously stable and trivial. An input signal $\hat \beta_{\IN}$ injected into the waveguide would scatter of as 
    \begin{align}
        \hat \beta_{\rm out}[\omega] = e^{i \phi[\omega]}  \hat \beta_{\rm in}[\omega]
        , \; \:
        \phi[\omega] = 2 \rm{arg} 
        \left[ i(\omega - \Lambda ) + \frac{\kappa}{2} \right],
    \end{align}
hence,  the input in the Bogoliubov basis is just reflected with a phase $\phi[\omega]$.
This phase is crucial to obtain any net amplification, which becomes obvious when considering the transformation back into the original frame
\begin{align} \label{Eq.DBAoutputCavBas}
\left(
\begin{array}{c}
 \hat  a_{\OUT}[\omega]
 \\
  \hat  a_{\OUT}^{\dag}[\omega] 
\end{array}
\right)
 =  
 \mathcal S^{-1}(r)
 \left[
\begin{array}{cc}
  e^{ + i  \phi[\omega] } & 0
 \\
 0 &  e^{- i  \phi[\omega] }
\end{array}
\right]
\mathcal S(r)
 \left(
\begin{array}{c}
 \hat  a_{\IN} [\omega]
 \\
  \hat  a_{\IN}^{\dag} [\omega]
\end{array}
\right) ,
\end{align}
with the single mode squeezing transformation $\mathcal S(r)$ 
  \begin{align}
\left(
\begin{array}{c}
  \hat \beta
 \\
  \hat \beta^{\dag} 
\end{array}
\right) 
 =   
\;
 \mathcal S(r)  \;
 \left(
\begin{array}{c}
 \hat  a
 \\
  \hat  a^{\dag} 
\end{array}
\right), 
\; 
 \mathcal S(r) = 
 \left(
\begin{array}{cc}
\cosh r & \sinh r
 \\
\sinh r & \cosh r
\end{array}
\right),
\end{align}
with $\mathcal S^{-1}(r) \mathcal S(r) =  \mathbb{1}$.
Crucially, the phase  $\phi[\omega]$ ensures that the squeezing transformations do not cancel each other out. This lack of cancellation results in net amplification of input signals. This becomes even more transparent when we transform into the quadrature basis   
\begin{align}\label{Eq.DBAoutputQuadBas}
 \left(
\begin{array}{c}
 \hat  X_{\OUT}[\omega]
 \\
  \hat  P_{\OUT}[\omega]
\end{array}
\right)
 =&
\left[ 
\begin{array}{cc}
    \cos    \phi[\omega]
&
   -  e^{-2r}  \sin \phi[\omega]
 \\
      e^{+2r} \sin \phi[\omega]
&     \cos    \phi[\omega]
\end{array}
\right]
 \left(
\begin{array}{c}
 \hat  X_{\IN} [\omega]
 \\
  \hat  P_{\IN} [\omega]
\end{array}
\right),
\end{align}
with $\hat X = (\hat a + \hat a^{\dag})/\sqrt{2}$ and  $\hat P = -i (\hat a - \hat a^{\dag})/\sqrt{2}$ as our orthogonal quadratures. It becomes apparent that for 
$\phi[\omega] = n \pi/2, n \in \mathbb{Z}^{*}$ the off-diagonal components, which mix the quadratures in the output, vanish and we obtain quantum-limited amplification of the X-quadrature.  
    
To specify the parameters required for an single-mode Bogoliubov amplifier, we take the following operational approach. We imagine having a fixed decay rate $\kappa$, while being able to freely adjust the detuning $\Delta$ and drive $\nu$. Changing both parameters is feasible in nearly all experimental platforms. We chose both such that the energy of the Bogoliubov mode matches the photonic loss rate
    \begin{align}\label{eq.MatchingODPA}
    \sqrt{\Delta^2-\nu^2} = \frac{\kappa}{2} ,
    \end{align} 
which coincides with the condition $\phi[0] = - \pi/2$ for the cavity resonance, 
cf. Eq.(\ref{Eq.DBAoutputQuadBas}) for $\omega=0$.    
The corresponding frequency dependent gain now reads
\begin{align}\label{eq:Gain_Optimal}
        \mathcal G[\omega]
        =
        \frac{(\frac{2\omega}{D'})^4+ \mathcal  G_0^{\prime}}{(\frac{2\omega}{D'})^4+1},
        \hspace{0.5cm}
 \mathcal  G_0^{\prime} = e^{4r} =
 \frac{\Delta + \nu}{\Delta - \nu}  ,    
\end{align}
where $D' = \sqrt{2}\kappa $ is the effective bandwidth of the optimally detuned Bogoliubov amplifier (ODBA). While superficially similar looking to the gain profile in Eq.~(\ref{eq:Gain_Normal}) of a conventional DPA , the ODBA offers two distinct advantages. The first is that there is no gain-bandwidth limitation: the gain can be as large as desired without sacrificing bandwidth (see Fig.~\ref{Figure2}).   
The fact that the gain and bandwidth are independent  is of enormous utility in experiments. The other distinct advantage of the ODBA is the relative flatness of the gain profile around zero frequency. In a standard DPA, the gain is only flat around an extremely narrow range of frequencies $\omega$ which satisfy  $\omega \ll D$. In contrast, the gain of a ODBA is nearly constant around for frequencies $\omega$ near zero, with small leading order correction of the order $(\omega/D')^4$ (see Fig.~\ref{Figure2})
    
The ODBA still maintains one of the main attractive features of a conventional DPA: it can be used as a quantum-limited amplifier without any added noise. Such amplifiers are required for several tasks related to quantum computation and communication. Our scheme is relevant in several experiment platforms, such as optical, microwave and mechanical setups. It is especially well suited for Josephson amplifiers used to readout superconducting quantum circuits. Finally, we stress that the ODBA is simple to implement experimentally. It does not require additional hardware, but relies instead on a  carefully chosen choice of detuning and drive strength, both of which can be easily tuned in experiments.

    
\section{An ideal two-port squeezer: the two-mode Bogoliubov amplifier}

The previous section showed how exploiting dynamics that considerved the number of squeezed photons (Bogoliubov excitations) in a simple parametrically driven cavity was enough to realize a quantum amplifier with exceptional properties.  Here, we show how this basic idea becomes even more powerful in the setting of a two-cavity system.  

\subsection{The Optimally Imbalanced Parametric Amplifier: The OIBA}\label{SecOptImbalnecedDPA}

Parametric modulation of the coupling between two coupled cavity modes results in two basic interactions at the mean-field level: 
frequency conversion for modulating at the frequency difference of the mode pair,
or parametric amplification if one drives at the sum of their frequencies. Our two-mode Bogoliubov amplifier utilizes resonant versions of these interactions simultaneously.  We thus start with the mean-field Hamiltonian (rotating at the respective mode resonant frequencies)
 \begin{align} \label{Eq.:HamiltonianDbasis}
 \hH  =&        
   G_{1}  \hat d_1^{\dag}\hat d_2^{\dag} +   G_{1}^{\ast} \hat d_1^{\PD} \hat d_2^{\PD}    
+  G_{2}  \hat d_1^{\dag}\hat d_2       +   G_{2}^{\ast} \hat d_1^{\PD} \hat  d_2^{\dag}  .
 \end{align}
Here $\hat {d}_{n} (n\in 1,2)$ denotes the annihilation operator of mode $n$, and the coupling
coefficients $G_n$ contain the amplitude and the phase of two external modulation tones. 
With the latter we can control which quadratures of the modes are involved
in the interaction. 

\begin{figure} 
  \centering\includegraphics[width=0.5\textwidth]{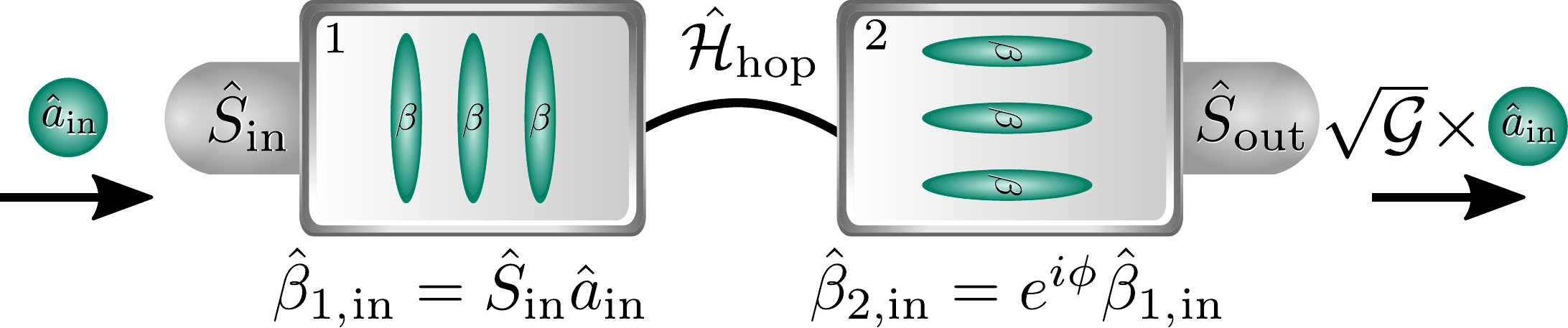}
 	\caption{
 	Illustration of the operation mode of the OIBA. An input signal $\hat a_{\IN}$ on mode 1 gets squeezed $(\hat S_{\rm in})$, converted to a mode 2 Bogoliubov quasiparticle and undergoes a second squeezing transformation before leaving mode 2. Crucially, due to the conversion process between the modes the squeezing transformations do not cancel each other out and the output signal is amplified.
 	}
 	\label{Figure3}
\end{figure}

Unlike the amplifier of Sec. \ref{SecOptDetunedDPA}, there is no explicit detuning term here.  Nonetheless, our system can be made dynamically stable (even without any external dissipation) by constraining $G_1$ to be smaller than $G_2$.  In this regime, the intracavity dynamics of our system is yet again simply understood in terms of Bogoliubov modes whose total excitation number is conserved.  Defining the squeezing parameter $r$ via $\tanh 2 r = G_{1}/G_2$, and introducing canonical Bogoliubov-mode lowering operators $\hat \beta_n =   \hat  d_{n}\cosh r +  \hat d_{n}^{\dag} \sinh r $,  Eq.~(\ref{Eq.:HamiltonianDbasis}) 
takes the form 
\begin{align}\label{Eq.BogHam}
 \hH_{\HOP} =& \; \widetilde G \;   \hat \beta_1^{\dag} \hat  \beta_2 + h.c. ,
  \hspace{0.3cm}
 \widetilde G  = \sqrt{G_{2}^2 - G_{1}^2}.
\end{align} 
Eq.~(\ref{Eq.BogHam}) describes a simple hopping interaction that converts excitations between the local Bogoliubov modes at a frequency $\tilde{G}$.  At first glance, this might seem to be of zero utility for amplification.  This however misses the fact that when we now couple modes $1$ and $2$ to input-output transmission lines or waveguides, excitations enter and the leave the system as photons, not as Bogoliubov excitations.  This then provides a simple heuristic picture for our amplification mechanism as illustrated in Fig.\ref{Figure3}:  Input photons on mode 1 are converted to $\beta_1$-mode excitations; this involves a single-mode squeezing transformation $\hat S_{\rm in}$. The hopping interaction in Eq.~(\ref{Eq.BogHam}) then converts the excitations to $\beta_2$ with a phase shift. The excitations in $\beta_2$ are then converts to photons in the transmission line coupled to mode 2; this involves a second single-mode squeezing transformation $\hat S_{\rm out}$.

The above heuristic picture is borne out by an explicit calculation of the system's scattering matrix. Without loss of generality we take the coefficients $G_1, G_2$ to be real, and couple both cavity modes to external waveguides with symmetric coupling strengths $\kappa_{n} = \kappa$ (see Appendix \ref{AppAsymmetries} for the asymmetric case). From the Langevin equations of the system we find the eigenvalues of the dynamics 
\begin{align}\label{Eq.Eigenvalues}
 \epsilon_{1,2} = - \frac{\kappa}{2} \pm i \widetilde G , \hspace{0.5cm} \widetilde G  \equiv \sqrt{ G_{2}^{2} - G_{1}^2},
\end{align} 
we see that if we choose $\tilde{G} = \kappa/2$, i.e., for optimally imbalanced interaction strengths $G_{1,2}$ in Eq.(\ref{Eq.:HamiltonianDbasis}), we have a situation where the splitting of the system's normal modes is exactly equal to their width.  As in Sec. \ref{SecOptDetunedDPA}, this matched-splitting operation leads to a number of exceptional properties, including an extremely flat gain versus frequency profile.  We thus follow on this special tuning point in what follows. To note is, that an amplifier with no gain-bandwidth limit is as well obtained for a balanced choice of coupling strengths (with the price of a lower bandwidth), please see Appendix \ref{app_QND} for details.

The output for $\tilde{G} = \kappa/2$ yields in the Bogoliubov basis 
\begin{align}  \label{Eq.ScattIBAgeneral}
    &
\left[
\begin{array}{c}
  \hat \beta_{1,\OUT}[\omega] 
 \\
  \hat \beta_{2,\OUT}[\omega] 
\end{array}
\right]
 =   e^{i\phi[\omega]} \mathcal P^{-1}
 \left[
\begin{array}{cc}
 \cos \theta[\omega] &   \sin \theta[\omega]
 \\
 - \sin \theta[\omega]  &  \cos \theta[\omega]
\end{array}
\right] 
\mathcal P
 \left[
\begin{array}{c}
\hat \beta_{1,\IN}[\omega]  
 \\
\hat \beta_{2,\IN}[\omega]  
\end{array}
\right]. 
\end{align}
The matrix operations here acting on the Bogoliubov input modes here have a simple interpretation.  They
correspond to the input excitation passing through a sequence of phase shifters.  The corresponding frequency dependent beam-splitter angle, and phase shifter matrix are: 
\begin{align}
    \theta[\omega] =  \textup{arcsin} 
    \left[ \frac{4\omega^4}{\kappa^4}+ 1  \right]^{-\frac{1}{2} },
\mathcal P = 
\left( 
\begin{array}{cc}
- i & 0 
 \\
0 & 1
\end{array}
\right)   . 
\end{align} 
In addition we have the frequency-dependent phase shift 
\begin{align} 
    \phi[\omega] = \textup{atan} 
    \left[
    \frac{2 \frac{\omega}{\kappa} }{1 - \frac{2\omega^2}{\kappa^2}  }
    \right] .
\end{align} 
The output of the mode $i=1,2$ contains now contributions from both modes, in contrast to the single-mode ODBA, where the input signal is simply reflected with a phase. This phase was the important ingredient to avoid the cancellation of the squeezing transformation,  cf. Eq.~{\ref{Eq.DBAoutputCavBas}}. 
Here the situation is slightly different:  to prevent the squeezing operations from cancelling (hence generating amplification), the beam-splitter operation is now crucial.  
To see this we can consider the scattering behavior in the original mode-basis.   
and work in the basis of the orthogonal quadratures
$ \hat X_{n} =   (\hat  d_{n}^{\PD}  +  \hat d_{n}^{\dag} )/\sqrt{2}$
and
$\hat  P_{n}  = - i (\hat  d_{n}^{\PD}  -  \hat d_{n}^{\dag})/\sqrt{2}$.
The scattering matrix relating input and output fields becomes  $(\textbf{X}_{\OUT}[\omega] = \textbf{s}[\omega] \textbf{X}_{\IN}[\omega] )$  
 \begin{align}  \label{ScatteringOIBA}
 \textbf{s}[\omega] 
 =&    e^{ i\phi[\omega]} \;
 \mathcal S_{Q}^{-1}(r) \;
 \mathcal B(\theta[\omega])
 \;
 \mathcal S_{Q}(r)
  \nonumber \\
  =& \frac{e^{ i\phi[\omega]} }
             {\sqrt{\frac{4\omega^4}{\kappa^4}+ 1 }  } 
             \left(
\begin{array}{cccc}
 \cot \theta[\omega]  & 0 & 0 & -e^{-2 r}   \\
 0 & \cot \theta[\omega]  & e^{2 r}  & 0 \\
 0 & -e^{-2 r}    & \cot \theta[\omega]  & 0 \\
 e^{2 r}   & 0 & 0 & \cot \theta[\omega]  \\
\end{array}
\right), 
\end{align}
in  the basis $\textbf{X} =[\hat  X_{1}, \hat  P_{1}, \hat X_{2}, \hat P_{2}]^T$,
and with the beam-splitter matrix
\begin{align}
 \mathcal B(\theta[\omega]) =&
  \left[
\begin{array}{cccc}
   \cos \theta[\omega]  &  0  &
  \sin \theta[\omega]  &  0
 \\
 0 &    \cos \theta[\omega]  & 
 0 &  \sin \theta[\omega] 
 \\
 - \sin \theta[\omega]  &  0  &
   \cos \theta[\omega]  &  0
  \\
 0 & -  \sin \theta[\omega]  &
 0 &    \cos \theta[\omega]
\end{array}
\right] ,
\end{align}
and the squeezing transformation for the quadrature basis (absorbing the phase shift $\mathcal P$)
\begin{align}
\mathcal S_{Q}(r) = \frac{1}{\sqrt{2}}
 \left[
\begin{array}{cccc}
  - i e^{+r} &    e^{-r} & 0 & 0 
  \\
 i e^{+r} &  e^{-r} & 0 & 0 
 \\
 0 & 0 &  e^{+r}  &  i e^{-r} 
 \\
 0 & 0 &  e^{+r} & - i e^{-r}   
\end{array}
\right],
\end{align} 
which describes local squeezing transformations. From the scattering matrix given in Eq.(\ref{ScatteringOIBA}) it becomes clear that the beam-splitter prevents the cancellation of the squeezing transformations, with the net result being phase-sensitive amplification involving a frequency conversion process.

We already see the remarkable features of the amplifier: 
the reflections (diagonal elements) are zero on resonance $(\theta[0] = \pi/2)$ and thus the system is perfectly impedance matched to its input ports. Moreover, the output of cavity 1 (2) contains
the amplified $P$-quadrature and squeezed  $X$-quadrature of cavity 2 (1),
allowing for a separation of input and output ports for the signal. 
In contrast to single mode-squeezing, i.e., realized via the interaction $\mathcal H_{s}  = \lambda \hat   d^{\dag} \hat   d^{\dag} + h.c.$, the squeezing/amplification here is also accompanied by a frequency conversion. Despite the fact that we use two degrees of freedom here (which might seem extraneous), we find that the amplification process is quantum-limited: it reaches the quantum-limit for phase-sensitive amplifiers of zero added noise \cite{Caves1982}. 
   
\begin{figure} 
  \centering\includegraphics[width=0.9\columnwidth]{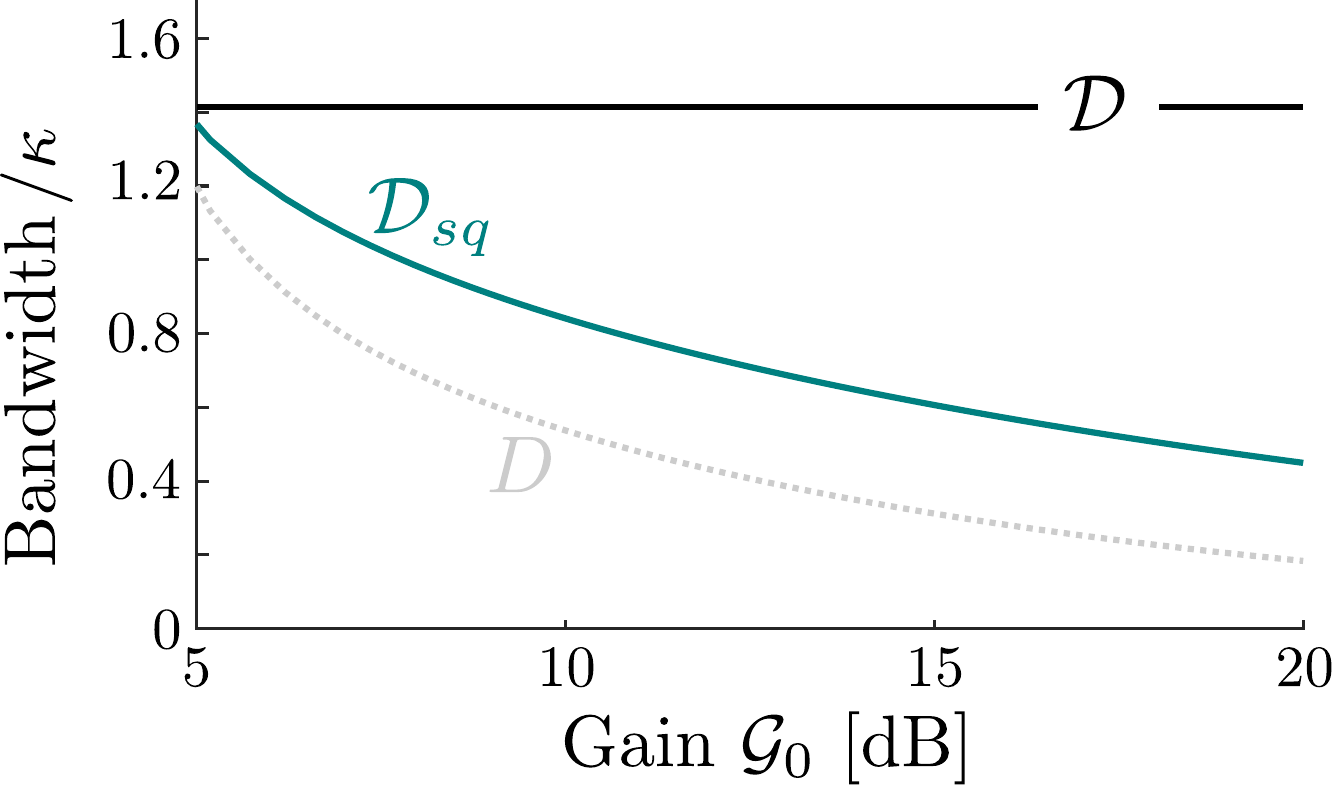}  
 	 \caption{
 	 Bandwidths for the  OIBA as a function of the gain $\mathcal G_{0}$. The black solid line depicts the amplification bandwidth $\mathcal D = \sqrt{2} \kappa$, which scales independently of the gain. This behavior does not translate to the squeezing bandwidth $\mathcal D_{sq}$ (emerald solid line), which decreases for increasing gain. However, the $\mathcal D_{sq}$ is enhanced compared to the squeezing bandwidth $ D \simeq   2\kappa/ \sqrt{\mathcal G_0}$ for the standard single mode setup (dotted grey line).  
 	\label{Figure4}
 	}
\end{figure}

Another crucial aspect of the optimally imbalanced Bogoliubov amplifier (OIBA) is the off-resonant gain behavior. Defining $ \mathcal G_0  = e^{4r}$ as the zero-frequency power gain, we find for the gain as a function of frequency
\begin{align}
  \mathcal G[\omega] \equiv   |s_{23} [\omega] |^2 =
                        \frac{ \mathcal G_0  }{  1 + \left[ \frac{2 \omega }
                             {  \mathcal D }\right]^4 } ,
                        \hspace{0.2cm}
                        \mathcal D = \sqrt{2} \kappa, 
\end{align}
thus, the gain scales linearly with the (tunable) coupling strengths $G_{1,2}$,
and the bandwidth over which amplification is possible is not affected by the amount of the 
gain, i.e., this amplifier is not limited by a fixed gain-bandwidth product. The resulting bandwidth is $\mathcal D  =   \sqrt{2}  \kappa $, see Fig.~\ref{Figure4}. To note is, that via the inclusion of additional parametric processes with a damped auxiliary mode, the OIBA can as well be rendered nonreciprocal, i.e., that it amplifies signals unidirectionally while still possessing the extraordinary features of the reciprocal OIBA, see \ref{AppNonreciprocal} for details.

An additional important question is whether this remarkable feature of a gain-independent bandwidth also manifests itself in the output squeezing.  As discussed in Sec.~\ref{SecRecapDPA},  a conventional single-mode squeezer has an amplification bandwidth scaling with
$ D \simeq   2\kappa/ \sqrt{\mathcal G_0}$, which coincides with the squeezing bandwidth. 
This squeezing bandwidth is defined as the frequency range over which the amount of squeezing is within $3$~dB of the maximal on-resonance value.  
To determine the squeezing bandwidth for the Bogoliubov amplifier we consider the symmetrized output noise spectra of the $X_1$-quadrature 
\begin{align}
 \frac{ \bar S_{X_1 X_1} [\omega]}{\bar S_{\textup{SN}}} =&     
                          \frac{  1  }{  1 + \left[ \frac{2 \omega }{ \mathcal D}\right]^4     } 
                          \left( \left[ \frac{2 \omega }{ \mathcal D}\right]^4    
                          + \frac{1}{\mathcal G_0}     \right) .  
\end{align}  
with $ \bar S_{\textup{SN}}  = 1/2$ as the shot-noise value.
The first term describes vacuum fluctuation driving cavity 1, while the second term describes the squeezed cavity-2 noise.
As discussed above, on resonance, i.e., $\omega = 0$, and $\bar n_2^T = 0 $ the amount of squeezing scales inversely with the gain just as in a single-mode setup.
However, the noise contribution from mode-1 becomes relevant for finite frequency;
we find for the squeezing bandwidth $\mathcal D_{sq}  \simeq   \sqrt{2} \kappa  ( \mathcal G_0)^{-1/4}$. Hence the gain-independence of the amplification bandwidth does not translate to the squeezing bandwidth. However, it is notable that the squeezing bandwidth is enhanced compared to a single-mode setup, i.e., for 
the same gain value $\mathcal G_{0} = \mathcal G_{s} \equiv \mathcal G$ we find $\mathcal D_{sq}/D  \simeq  \mathcal G^{1/4}/\sqrt{2}$.  
Note, without cavity-1 noise contribution in the spectrum $ \bar S_{X_1 X_1} [\omega]$, the squeezing bandwidth would scale independently of the gain.

\subsection{OIBA: Experimental Results} \label{SecExpResults}

\begin{figure}
    \centering
    \includegraphics[width=0.5\textwidth]{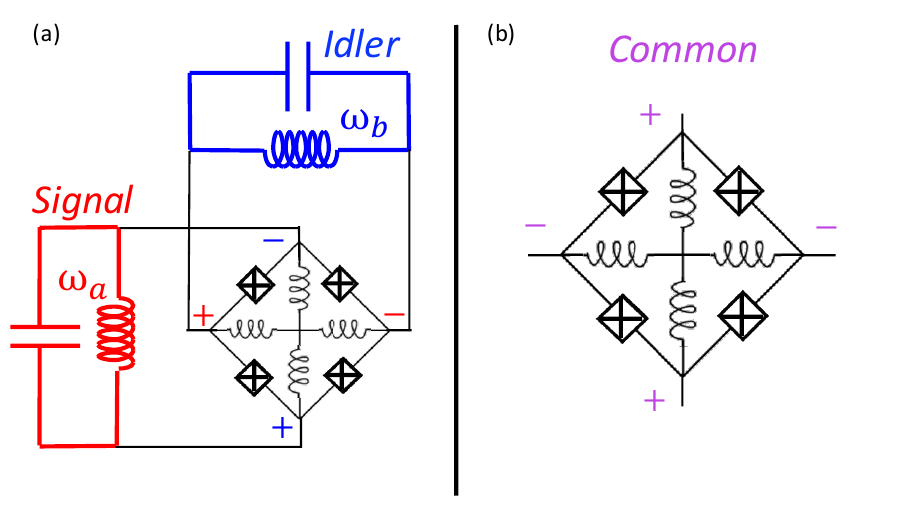}
    \caption[Signal (red), idler (blue), of a JPC. JPC] {{(a) Signal(red) and Idler(blue) modes of a JPC connected to differential excitations of the central JRM.(b) The common mode of the JPC and JRM connects to both external modes and is typically used for pumping rather than signal input and output.}}
    \label{Figure5}
\end{figure}

We demonstrate the power and functionality of these theoretical predictions using a shunted Josephson Parametric Converter (JPC) whose properties and fabrication are described in Ref.~\cite{Chien2019}. The core of this amplifier is a shunted Josephson Ring Modulator (JRM); a ring of four nominally identical Josephson junctions, shunted with linear inductors, as shown in Fig.~\ref{Figure5}. The ring is threaded with an external magnetic flux which creates three-wave mixing among the devices three modes.

The Hamiltonian of this  device written in terms of the normal modes is 
\begin{align}
\begin{split}
    H_{JRM} = & -4E_J [ \cos(\dfrac{\varphi_X}{2})\cos(\dfrac{\varphi_Y}{2})\cos(\varphi_Z)\cos(\dfrac{\varphi_{ext}}{4}) \\
     & + \sin(\dfrac{\varphi_X}{2})\sin(\dfrac{\varphi_Y}{2})\sin(\varphi_Z)\sin(\dfrac{\varphi_{ext}}{4})]  \\
    & +\dfrac{E_L}{4}(\varphi_X^2 +\varphi_Y^2 +2\varphi_Z),
    \end{split}
    \label{HRJM_full}
\end{align}
where $E_L= \dfrac{\Phi_0^2}{L}$. We refer to these modes as the signal, idler, and common mode of the JRM. These mode excitations are again noted in Fig~\ref{Figure5}.

The four Josephson junctions on the outer arms of the JRM provide nonlinear couplings between the eigenmodes of the circuit \cite{Abdo2011}. Assuming that the ground state of the circuit is $\varphi_X = \varphi_Y =\varphi_Z = 0$ and that it is stable as we tune the external magnetic flux bias, we can expand the nonlinear coupling terms around this ground state and make appropriate substitutions to rewrite the Hamiltonian in terms of the raising and lowering operators, up to 3rd order, as in \cite{Schackert2013, chenxu2019}
\begin{align}\label{3bodyHamiltonian}
\hH_{JPC} = 
\hH_{0}
+ g_3(\hat a + \hat a^\dag)(\hat b+ \hat b^\dag)(\hat c+ \hat c^\dag),
\end{align}
where $\hH_{0}$ denotes the Hamiltonian of the three uncoupled modes and the $g_3$ term represents the strength of the 3-wave mixing that gives rise to the gain and conversion processes in the JRM \cite{Sliwa2015}. These 3-wave processes are then driven with a far off-resonance stiff tone to create effective two-body interactions. Note that the expansion of the cosine potentials also results in higher order terms that have been neglected, as they have significantly smaller magnitude.  These terms will nonetheless hinder performance compared to the ideal 3-wave mixing Hamiltonian. 

To achieve the Optimally Imbalanced Bogoliubov Amplifier (OIBA) experimentally, we start with two pump tones (one each at the sum and difference frequencies of the signal and idler modes) which drive the conversion and gain processes of the JPC with identical coupling rates. From here, the conversion tone amplitude is increased so that a large dip in signal reflection becomes visible. Then, we switch the read-out scheme to transmission ,using a mixer to undo the amplifier's frequency conjugation so that we can compare same-frequency input and outputs in a Vector Network Analyzer. 

To increase the gain, ${\mathcal G}$, both pumps are both now individually increased slowly at the same rate. As long as the dip in reflection is still present in reflection, the correct ratio of the pumps is approximately maintained.

\begin{figure}
    \centering
    \includegraphics[width=0.5\textwidth]{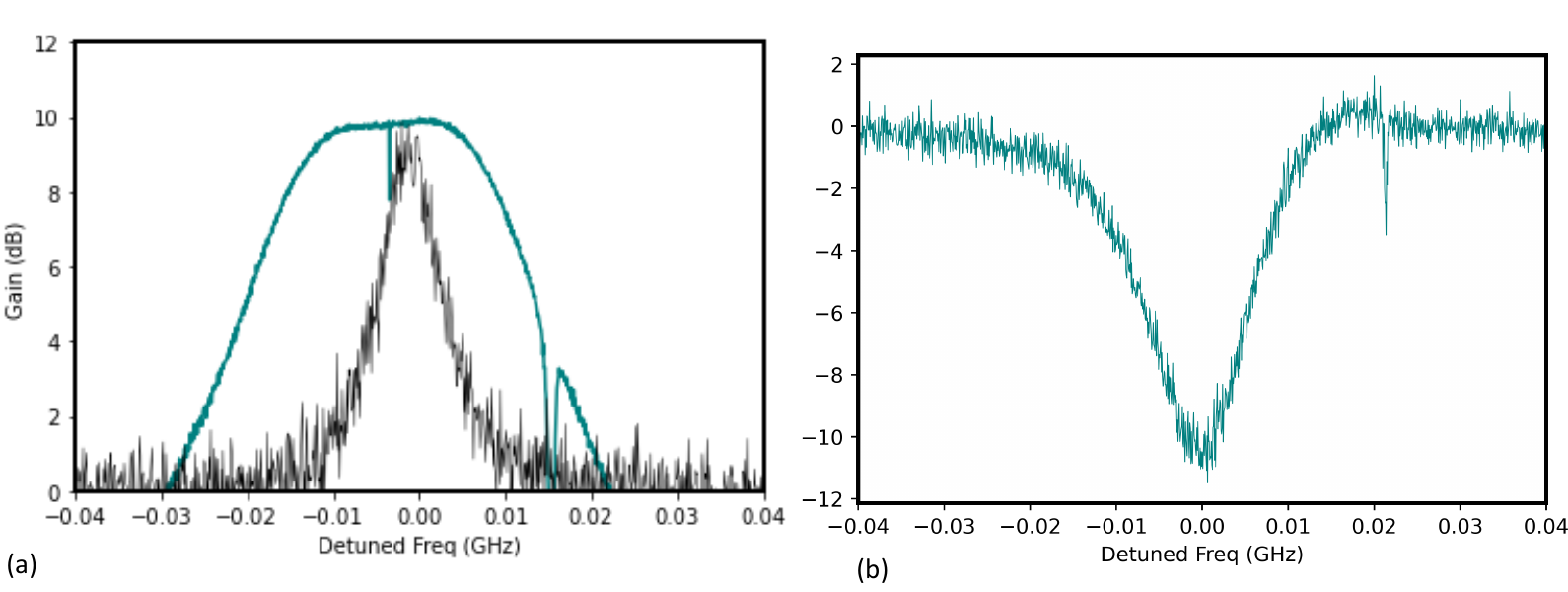}
    \caption{(a)OIBA-pumped amplifier (teal) transmission gain versus frequency, compared to a more standard amplification setup using a single pump tone (black); in both cases we have 10 dB of gain in reflection.  The OIBA scheme exhibits a dramatic improvement in bandwidth for the same amount of gain.(b) The reflection scattering parameter of the OIBA amplifier versus frequency, showing that there is almost perfect impedance matching on resonance (i.e.~almost vanishing reflection).}
    \label{Figure6}
\end{figure}

If we compare the response of the OIBA to 
a more standard non-degenerate amplification process using a single pump tone, we can see that both scattering parameters of the OIBA are superior. The $\approx10$~dB dip in Fig.~\ref{Figure6}~b. means that the amplifier is still approximately matched in reflection. The OIBA is still bi-directional, meaning that one can amplify both from the signal to the idler mode, but also in reverse. A fully directional amplifier is still necessary to achieve ideal practical operation, because stray photons can still end up traveling backwards down the amplification chain to the qubit \cite{Lecocq2017, Li2019}. Thus, in the OIBA measurements should still be performed with external isolators to protect the device being measured.

While the broad gain peak in transmission is not an ideal 20~dB, it has all the qualitative features predicted by theory, including the much larger bandwidth, flat peak, and phase-sensitive mode of amplification.  At this bias point the single gain pump achieves only about $5$ MHz bandwidth at this gain, which is typical for a standard Josephson Parametric Amplifier. The OIBA on the other hand, has a bandwidth of $\approx 33$ MHz. Theory predicts an even broader bandwidth for the OIBA,  but that rests upon having $\kappa_{sig} = \kappa_{idl}$, a condition that was slightly violated here ($\kappa_{sig} = 25$~MHz, $\kappa_{idl}= 20$~MHz). In addition, it was experimentally difficult to reach 20~dB of gain in this mode of amplification on this amplifier due to unwanted higher order terms. In the future, this difficulty could be alleviated by using a more ideal mixing element, such as an array of SNAILs, with suppressed higher-order terms. 

Despite these caveats, we stress that this experimental device still serves proving the validity of the basic concept.  The crucial observation is the ratio of bandwidths between the two kinds of amplifier modes:  the new OIBA scheme always yields a far larger bandwidth than the typical single-pump amplification setup (for the same external flux conditions). Further, the OIBA approach does not suffer from a gain-bandwidth product limit that constraints the standard single-pump approach to amplification.

\subsection{OIBA: Pump-Depletion}
%
%
We now show that the OIBA scheme has another strong advantage over more conventional approaches: it is more robust against pump depletion effects.
In the former section we have seen that OIBA amplifier can be realized in a superconducting-circuit setup which uses a JPC as the mixing element, and which is driven via two external microwave tones.
The relevant 3-wave mixing process is described via the Hamiltonian
\begin{align}
 \hH_{3W}  
	=&     \;  
		 g_1    \hat  a_{1}    \hat  d_{1}^{\dag}  \hat  d_{2}^{\dag}    
		+g_2    \hat   a_{2}    \hat d_{1}^{\dag}  \hat d_{2}    
		+ h.c. ,
\end{align}
here $\hat  a_{n}$ denote the pump modes which are driven strongly at frequencies $\omega_{P,1} = \omega_{1} + \omega_{2}$ and $\omega_{P,2} = \omega_{1} - \omega_{2}$
(and detuned from their resonance). Performing a displacement transformation  $\hat   a_{n} \rightarrow  a_{n}  e^{-i\omega_{P,n} t - i \varphi_{n}} + \delta \hat a_{n}$, we can decompose the pump modes into their classical amplitudes $a_{n}$ and corresponding fluctuations $ \delta \hat a_{n}$. 
The pump phases $\varphi_{n}$ will determine  which quadrature-quadrature coupling we obtain.
In general one assumes stiff pump modes, i.e., one sets $\hat  a_{n} = \ev{\hat   a_{n}}$ and neglects $\delta \hat   a_{n}$, then the resonant interaction simplifies to
Eq.~(\ref{Eq.:HamiltonianDbasis}) with $G_{n} = g_n \ev{\hat   a_n} e^{-i \varphi_n}$.

The stiff pump approximation breaks down for larger input signal power. Here backaction effects on the pump become relevant and limit the dynamical range of the parametric amplifier.
To go beyond the stiff pump approximation we derive the equations of motion for the pump modes' expectation values 
\begin{align}
 \frac{d}{dt} \ev{\hat a_{1}} =& - \frac{\gamma_1}{2} \ev{\hat  a_{1}} - \sqrt{\gamma_{1}} a_{1,\IN} - i g_1 \ev{ \hat  d_{1}   \hat  d_{2}}   ,
 \nonumber \\ 
 \frac{d}{dt} \ev{\hat a_{2}} =& - \frac{\gamma_2}{2} \ev{\hat  a_{2}} - \sqrt{\gamma_{2}} a_{2,\IN} - i g_2 \ev{ \hat d_{1}   \hat  d_{2}^{\dag}   }   ,
\end{align} 
where $\gamma_{1,2}$ denote the damping of the pump modes and $a_{1,2,\IN}$ correspond to the amplitude of the respective pump tone. 
The coupling to the correlators $\langle\hat d_{1}\hat  d_{2}^{(\dag)}\rangle$ in the equations for the pump modes describe the backaction effects, 
the latter would be neglected under a stiff pump approximation. We evaluate the correlators on a mean-field level \cite{Abdo2013} 
and define
\begin{align}
 \Sigma_{1,\EFF} \equiv & \;  i \frac{g_1}{\ev{a_1}} \ev{  \hat  d_{1} \hat  d_{2}   } =  \frac{\gamma_{1,\EFF}}{2} + i \Omega_{1,\EFF},
 \nonumber \\
  \Sigma_{2,\EFF} \equiv&  \;  i \frac{g_2}{\ev{a_2}} \ev{   d_{1} \hat  d_{2}^{\dag}    } =  \frac{\gamma_{2,\EFF}}{2} + i \Omega_{2,\EFF},
\end{align}
with the effective damping rates
\begin{align} \label{Eq.EffectiveDecay}
  \gamma_{1,\EFF}  =&  + \frac{\gamma_1}{2}  
    \frac{  \sqrt{\mathcal C_1} }{\left[1 +\mathcal C_{+} \mathcal C_{-} \right]^2} 
   \bigg\{    \frac{ X_{n,\IN}^2}{\bar n_{1,\IN}}  \mathcal C_{+}   
         -    \frac{P_{n,\IN}^2}{\bar n_{1,\IN}}   \mathcal C_{-}      
   \bigg\} 
   +   \gamma_{1,\EFF}^{\textup{vac}},
     \nonumber \\  
  \gamma_{2,\EFF}  =&  \mp \frac{\gamma_2}{2}  
    \frac{  \sqrt{\mathcal C_2} }{\left[1 + \mathcal C_{+} \mathcal C_{-}\right]^2} 
   \bigg\{    \frac{ X_{n,\IN}^2}{\bar n_{2,\IN}} \mathcal C_{+}  
         +    \frac{P_{n,\IN}^2}{\bar n_{2,\IN}}  \mathcal C_{-}    
   \bigg\}  , 
\end{align}
with $\mathcal C_{\pm} = \sqrt{\mathcal C_{2}} \pm \sqrt{\mathcal C_{1}}$,
the cooperativities $\mathcal C_{n} = 4 G_n^2/\kappa^2 $, and
where the minus sign in $\gamma_{2,\EFF} $ refers to an input in cavity 1.  
The damping rate associated with vacuum fluctuations driving mode-1 reads 
\begin{align}
 \gamma_{1,\EFF}^{\textup{vac}} = \frac{ 2 g_1^2}{\kappa}
     \frac{1 }{1 + \mathcal C_{+} \mathcal C_{-} },
\end{align}
which is negligible as it neither scales with the input signal nor the gain.
In addition, the frequency shifts yield
\begin{align}
      \Omega_{n^{\prime},\EFF}   =& (-1)^{n^{\prime}+1} \; \frac{\gamma_{n^{\prime}}}{2}  
    \frac{ \sqrt{\mathcal C_1 \mathcal C_2}   }{\left[1 + \mathcal C_{+} \mathcal C_{-}\right]^2} \; \frac{   X_{n,\IN} P_{n,\IN} }{\bar n_{n^{\prime},\IN}} , 
\end{align}
which become only relevant if we have an input signal simultaneously in both quadratures, i.e., $ X_{n,\IN}$ and $ P_{n,\IN}$. Note, determining all backaction effects requires a self-consistent calculation, i.e., the cooperativities appearing in the expression for the effective decay rates $\gamma_{n, \EFF}$ and the frequency shifts $\Omega_{n,\EFF}$ depend on the pump amplitude and vice versa. 

\begin{figure}[t] 
  \centering\includegraphics[width=0.5\textwidth]{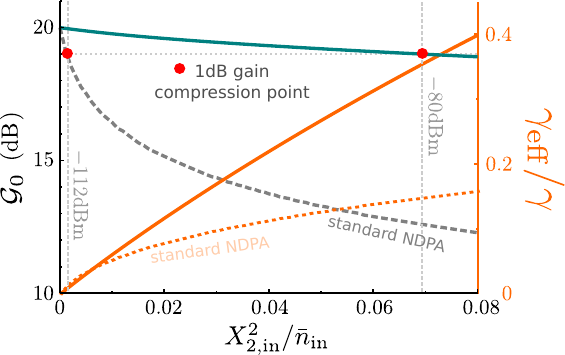}  
 	\caption{ 
 	Gain saturation pump-mode backaction effects in the OIBA amplifier, as a  function of the input signal strength $X_{2,\IN}^2/\bar n_{\IN}$,
 	where $\bar n_{\IN}$ denotes the average number of pump photons required to obtain $20$dB of gain. The emerald (orange) solid line depicted the gain (effective decay rate)  for the OIBA. 
 	For comparison the effective damping and the depleted gain for a standard parametric amplifier are plotted as well [grey dashed/orange dotted line]. All numerical results were obtained self-consistently.
 	Parameters are $\gamma_1/\kappa = \gamma_2/\kappa = 12$ and  $g_1/\kappa = g_2/\kappa = 0.014$. Taking typical JPC-parameters from Ref.~\cite{Abdo2013} the  $1$~dB compression point for the Bogoliubov amplifier is shifted by $32$~dBm  
 	in comparison to the standard paramp [explicit values denoted in graph]. 
 	}
 	\label{Figure7}
\end{figure}

Including the backaction onto the pump modes we have now pump amplitudes which depend as well on the phase of the input signal, a situation which deviates from  a standard single-tone (phase-insensitive) parametric amplifier. The latter case would be recovered by setting $\mathcal C_{2 } = 0$ in the upper expressions. 
Note that the OIBA requires a fine tuning of the pump amplitudes: for optimal performance the matching condition $\tilde G = \kappa/2$ has to be achieved.  Including pump depletion effects, one has a potential problem, as now, increases in the input signal strength could cause one to violate the matching condition. To quantify this effect we define 
\begin{align}
 \mathcal C_{1,2} (\gamma_{1,\EFF} ,\gamma_{2,\EFF} ) \equiv  
                         \mathcal C_{n, \EFF} 
                      =  \frac{ \mathcal C_{n} }{  ( 1 + \bar \gamma_{n,\EFF})^2    } \equiv \mathcal C_{n} \chi_{n},
\end{align}
as the effective cooperativities
with $\bar\gamma_{n,\EFF} =  \gamma_{n,\EFF} /\gamma_n $ and $\mathcal C_{n}$ denoting the undisturbed cooperativities, i.e.,
$\mathcal C_{n} = \mathcal C_{n,\EFF} $ for $\chi_{n} = 1$.  
The alteration of the cooperativities affects the matching condition for the OIBA amplifier, and we define the deviation from the optimal matching as  $\delta\mathcal C = \mathcal C_{2,\EFF} - \mathcal C_{1,\EFF} - 1$.
Assuming $\chi_n\approx \chi$ the deviation can be approximated as 
$\delta\mathcal C \approx \chi -1 $ and thus can be assumed to be small .
The effect of this mismatch is a minor back-reflection of the input signal,  
e.g. for the example in Fig.~\ref{Figure7} around $0.04\%$ of the signal are reflected at the $1$dB compression point of the amplifier. 

To analyze the back-action effects further we consider the situation where an input is injected in the X-quadrature of the second cavity.
The effective decay rates in Eq.~\ref{Eq.EffectiveDecay} can be approximated as   
\begin{align}\label{Eq.EffectiveDecayApprox}
  \gamma_{n,\EFF}   
    \approx 
   \frac{\gamma_n}{16}  \mathcal G_{\EFF} \frac{ X_{2,\IN}^2}{\bar n_{n,\IN}}  , 
\end{align}
thus, the effective decay rates scale linearly with the effective power gain $\mathcal G_{\EFF}$ and the input signal. This coincides with the scaling for the standard parametric amplifier (for $\mathcal C_{2 } = 0$ )\cite{Abdo2013},  leading to gain saturation  if the input signal strength is increased for constant gain. In a standard parametric amplifier the effective gain saturation scales as
\begin{align}
 \sqrt{ \frac{ \mathcal G_{PA,\EFF} }{\mathcal G_{PA}} } =&  
                                 \frac{1}{\sqrt{\mathcal G_{PA}}}\frac{    \chi^{-1} + \mathcal C  }{  \chi^{-1} - \mathcal C  } 
                                \approx
                                   1  -     \sqrt{\mathcal G_{PA}}   \bar \gamma_{ \EFF},
\end{align}
with $\mathcal G_{PA} = (1 + \mathcal C)^2/(1 - \mathcal C)^2 $ and the approximation in the second step holds for small effective decay rates. Crucially, the reduction of the gain is enhanced by the amplitude gain in this standard parametric amplifier.  

Taking that the OIBA effective decay rates in Eq.(\ref{Eq.EffectiveDecayApprox}) scale linearly with the effective gain, we should at first side expect similar saturation effects in the OIBA. 
The effective decay rates are even enhanced compared to the standard case, see Fig.~\ref{Figure7}. 
Importantly, the enhanced decay rates do not mean that the gain saturates earlier. In contrast, the effective decay rates are enhanced because the effective gain is larger, see Fig.~\ref{Figure7}. This originates in the modified scaling of the gain with the pump amplitude. The OIBAs effective gain saturation scales as  
\begin{align}
 \sqrt{ \frac{ \mathcal G_{0,\EFF} }{\mathcal G_{0}} } =&  
        \frac{ 2    \sqrt{  \chi }  }{   \chi    +  1 }  
         \approx 
         1 - \frac{ \bar\gamma_{n,\EFF}^2 }{2}  ,
\end{align}
where we approximated $\chi_{n}  \approx \chi$ and expanded the resulting expression for small effective decay rates. Clearly, this is a much favorable scaling, as the reduction of the gain is not enhanced by a gain-factor.  
This explains why the saturation of the OIPA sets in for much higher input signal strength,  cf. Fig.~\ref{Figure7}. This robustness should hold true for similar amplifiers without a gain-bandwidth product,  where the gain scales directly with the pump amplitude.


\section{The class of Bogoliubov amplifiers}
%

The previous sections have established a general principle for realizing amplification without instability, by using stable Hamiltonians in the Bogoliubov basis.  We now show that these ideas can be further generalized to a wide class of multi-mode systems.   
We consider $N$ Bogoliubov modes which obey the stable dynamics   
 \begin{align}\label{eq.GenBogDef}
  \hH =   \sum_{i,j = 1}^{N} \; \lambda_{i,j} \; \hat  \beta^{\dag}_{i}  \hat  \beta_{j}   ,  
 \end{align}
 with coupling strength $\lambda_{i,j} $. 
The Bogoliubov modes $\hat \beta_{n}  $ 
are obtained via the general squeezing transformation 
 \begin{align}
  \hat S_n = e^{ R_{n,m}\left( \hat  a_{n}\hat   a_{m} -  \hat   a_{n}^{\dag}  \hat a_{m}^{\dag} \right)}, 
  \; \;
  R_{n,m} = \frac{r_{n}}{1 +\delta_{n,m}} ,
  \end{align}
acting on the cavity mode operators $\hat a_{n,m} $ in the unsqueezed basis, i.e.,
$\hat \beta_{n} = \hat S^{\dag}_n \hat a_{n} \hat S_{n}$.  
Here $\delta_{n,m} $ denotes the Kronecker delta.
The squeezing transformation corresponds to either single-mode squeezing ($n=m$) or two-mode squeezing ($n \neq m$). The squeezing parameter $r_n$ depends on 
parameters in the respective unsqueezed cavity basis and are specified case by case.
We can consider now two different classes containing a phase-sensitive and a phase-insensitive version each: namely the class of detuned Bogoliubov amplifiers and the class of imbalanced Bogoliubov amplifiers.

\subsubsection{Detuned Bogoliubov amplifiers}
We start with the detuned Bogoliubov amplifiers, they are obtained by setting
$\lambda_{ij} = \delta_{ij} \lambda$ in  Eq.~(\ref{eq.GenBogDef}), so that
the Hamiltonian in the Bogoliubov basis reduces to
 \begin{align}
  \hH =   \lambda  \; \sum_{i=1}^{N} \; \hat   \beta_{i}^{\dag}   \hat   \beta_{i} .
 \end{align}
For $n=m$, i.e., single-mode squeezing, and $N =1$ we have only a single Bogoliubov mode
and recover the ODBA discussed in Sec \ref{SecOptDetunedDPA}. This means that by setting $\lambda =   \sqrt{\Delta^2 - \nu^2}$  
we obtain the phase-sensitive amplifier described by the Hamiltonian in Eq.~(\ref{eq:DPA}) 
in the unsqueezed cavity basis. However, it is also possible to design a phase-insensitive version via a two-mode
squeezing transformation, i.e., for  $n \neq m$ and the two Bogoliubov modes $\hat \beta_{n} = \cosh r \; \hat   a_{n}  + \sinh r \; \hat   a_{m}^{\dag}$ with $n,m = 1,2$. 
The squeezing parameter yields then $\tanh 2r = G/\Delta$, while the Bogoliubov mode energy becomes $\lambda = \sqrt{\Delta^2 - G^2}$. 
Here the detuning $\Delta$ and the two-mode squeezing strength $G$ are defined in the original basis as
  \begin{align}
   \hH =    \Delta \left( \hat  a_{1}^{\dag} \hat a_{1} + \hat  a_{2}^{\dag} \hat  a_{2} \right)  
         +  G      \left[ \hat  a_{1}^{\dag} \hat   a_{2}^{\dag} + \hat  a_{1} \hat   a_{2}  \right] .
  \end{align}
Thus we simply have a detuned two-mode squeezing interaction among two cavity modes,
in analogy to the ODBA, which involves a detuned single-mode squeezing interaction. 
Note that both kinds of detuned Bogoliubov amplifiers realize amplification without instability when the energy of the Bogoliubov mode matches the photonic loss rate,  i.e., $\lambda  = \kappa/2$ as done in Eq.~(\ref{eq.MatchingODPA}). Thus, by simply detuning the standard single or two-mode squeezing interactions the amplification process is stabilized and the resulting bandwidth is independent of the gain. 

\subsubsection{Imbalanced Bogoliubov amplifiers}

A second approach to achieving amplification using stable dynamics is to have a Hamiltonian that describes hopping interactions between localized Bogliubov modes.  
Considering the simplest case of two modes and $i \neq  j$ in Eq.~(\ref{eq.GenBogDef}) the Hamiltonian in the Bogoliubov base becomes
 \begin{align}
  \hH =  \lambda \; \left( \hat  \beta^{\dag}_{1}  \hat \beta_{2}  +  \hat  \beta_{1}  \hat  \beta_{2}^{\dag} \right),
 \end{align} 
which corresponds to a swapping of excitation between the modes,  
i.e., the number of Bogoliubov quasiparticles is conserved and they coherently oscillate back and forth between the modes $\beta_{1}$ and $\beta_{2}$.
Based on this stable dynamics in the Bogoliubov basis we can repeat our protocol to
obtain phase-sensitive ($n=m$) and phase-insensitive amplification  ($n \neq m$) without instability. 

\begin{figure} 
  \centering\includegraphics[width=1.0\columnwidth]{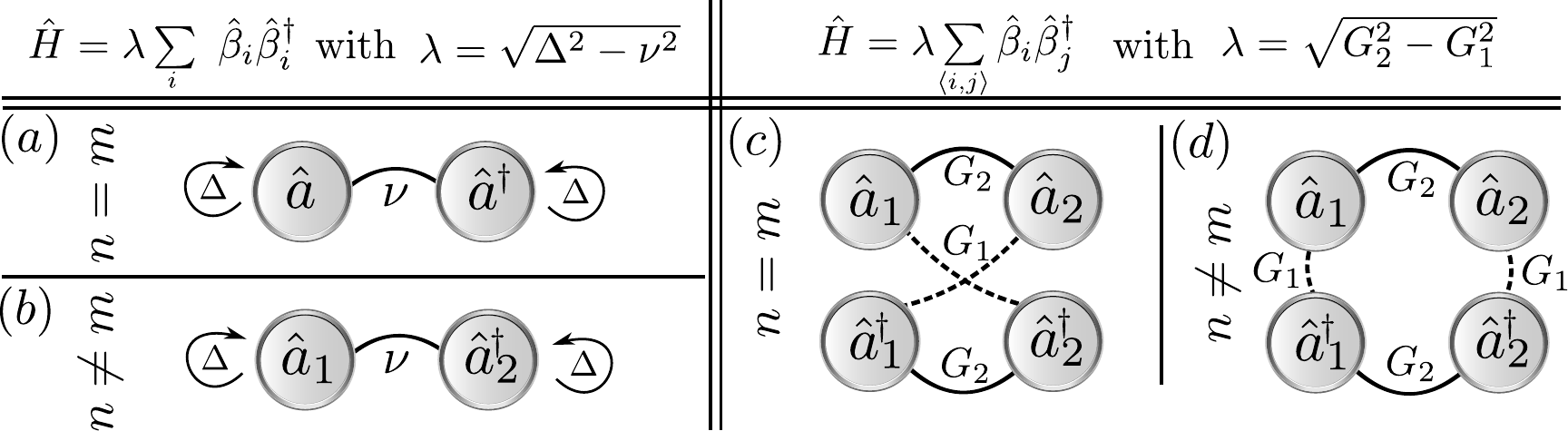}  
 	 \caption{ 
 	 The class of one and two-mode Bogoliubov amplifiers in the unsqueezed cavity basis. Configurations (a,c) realize phase-sensitive amplification without an instability,
 	 while the configurations (b,d) correspond to the phase-insensitive counterpart. In sketch (c,d) $G_{1}$ denotes single- or two-mode squeezing in (d) and (c) respectively, while $G_{2}$ corresponds to a hopping interaction. 
 	 }
 	\label{Figure8}
\end{figure}

Figure \ref{Figure8}(c,d) depict sketches of the required configurations of the imbalanced Bogoliubov amplifiers in the unsqueezed basis. We find that in addition to single- or two-mode squeezing interactions with strength $G_{1}$, we require a hopping interaction between the cavity modes $1$ and $2$ associated with strength $G_{2}$ in this original basis.
The sweet spot of operating without an instability is obtained for $\lambda = \sqrt{G_{2}^2 - G_{1}^2} = \kappa/2 $. Hence we refer to this class as the imbalanced Bogoliubov amplifiers, as the interaction strengths of the involved processes have to be imbalanced to match this condition.
Note that the bosonic Kitaev chain amplifier introduced in \cite{McDonaldPRX2018} can be viewed as a multi-mode realization of this kind of amplifier.

\section{Conclusion}
%
We have presented a novel class of quantum-limited amplifiers which operate effectively detuned from any instability. This mode of operation brings in the remarkable features of no gain-bandwidth limitation and a very flat frequency-gain profile. We showed that the removal of the instability is best understood in the basis of Boguliobov modes undergoing stable dynamics. Crucially, the transformations of an input signal into and out of the Bogoliubov basis are distinct and do not cancel each other out, leading to net amplification of the input signal. A theoretical analysis on the level of a mean-field ansatz shows that such  Boguliobov amplifiers are potentially more robust to detrimental backaction effects induced by large input signals. We introduced in detail the optimally imbalanced Bogoliubov amplifier (OIBA), which is based on a imbalanced frequency conversion and parametric amplification process. The OIBA, for which we presented proof-of principle experimental results, is an amplifier operating in transmission and is perfectly impedanced matched to its input ports.  These features make it an interesting candidate for a cascaded amplifier architecture, and for further applications to quantum signal processing. 

%
\section{Acknowledgements}
\label{sec_thanks}
%
AM acknowledges funding by the Deutsche Forschungsgemeinschaft through the Emmy Noether program (Grant No.~ME 4863/1-1)
and the project CRC 910.  AAC and AM acknowledge support from the Air Force Office of Scientific Research under Award No.~FA9550-19-1-0362, and the Army Research Office under Grant No.~W911NF-19-1-0328. OL, TZC, and MJH acknowledge support from the Army Research Office under Grant No.~W911NF-18-1-0144 and the National Science Foundation under Grant No. PIRE-1743717.

\newpage

\appendix
\section{Squeezing properties}
\label{app_Squee}

We can parametrize the squeezing obtained in the Bogoliubov amplifier 
via the squeezing parameter $r$, defined via the ratio $ e^{-2 r} = \bar S_{X_n X_n} [0]/ \bar S_{\textup{SN}}    $ 
with $ \bar S_{\textup{SN}}  = 1/2$ as the shot-noise value
and $\bar S_{X_n X_n} [0]$ denotes the symmetrized output noise of the $X_n$-quadratures. 
Here both quadratures, $X_{1}$ and $X_2$, are squeezed symmetrically hence we use only one squeezing parameter $r$, with $e^{-2 r} = 1/\mathcal G_0  $ for a pure vacuum input.
For the latter case the squeezed output light is pure. 
This changes once an input in mode-$n$ is accompanied by thermal fluctuations
originating from a bath with average occupation $\bar n_n^T$. 
For example, considering an input on cavity-2, the impurity of the output light of cavity-1 can be quantified by an effective thermal occupancy
\begin{align}
\bar  n_{\EFF}   =& \sqrt{ \bar S_{X_1 X_1} [0]  \bar S_{P_1 P_1} [0]} - \frac{1}{2} 
=  \; \bar n_2^T , 
\end{align} 
i.e., the output light is just as impure as the input light, which coincides with the situation for a single-mode squeezing setup. 
%
\section{Equally balanced amplifier}
\label{app_QND}
The Bogoliubov amplifiers require the optimal matching of the parametric coupling strength (e.g. OIBA) and the detuning (e.g. ODBA) to the coupling to the output port. However, obtaining an amplifier with no-gain bandwidth limitation is as well possible for a range of other parameters (with the price of a slightly reduced bandwidth). The onset of the this parameter region appears for a balanced choice of coupling strengths and/or detuning. We illustrates this briefly in what follows.

We start with a balanced choice of coupling parameters in the Hamiltonian given in Eq.~(\ref{Eq.:HamiltonianDbasis}), i.e., $G_{1} = G_{2} =  G  \in \mathbb{R}$, here
we obtain a quantum nondemolition (QND) interaction \cite{Braginsky1980}, i.e., both X-quadratures commute with the resulting Hamiltonian:
\begin{align}\label{Eq.:HamiltonianQND}
 \hH_{\QND} = 2 G \hat X_1 \hat  X_2.
\end{align}  
Such a process corresponds to a non-disturbing measurement
of the X-quadratures of each mode, where the P-quadratures obtain 
the information of the X-quadratures,  which themselves are not affected by the measurement. 
The combined balanced gain and conversion process realize a phase-sensitive amplifier, whose nonreciprocal version was briefly introduced in our 
former work \cite{Metelmann2015}. Crucially, as the X-quadratures are not affected by the QND-interaction no squeezing of an input signal is possible
and the amplifier cannot be impedanced matched on it's input port. 
 
The scattering matrix in the quadrature basis $\textbf{X} =[\hat  X_{1}, \hat  P_{1}, \hat  X_{2}, \hat P_{2}]^T$
becomes
\begin{align}
\textbf{s}[\omega] = 
e^{i\phi}  \left(
\begin{array}{cccc}
-  1    & 0 & 0 & 0 
\\
 0 & -  1 & \sqrt{ \mathcal G [\omega]} & 0 
 \\
 0 & 0 & -   1 & 0 
 \\
 \sqrt{ \mathcal G [\omega]} & 0 & 0 & -  1 
 \\
\end{array}
\right),
\hspace{0.1cm}
\mathcal G [\omega] = \frac{ \mathcal G_{Q}   }{\left[1 + \frac{ 4 \omega^2}{\kappa^2} \right]^2   } 
\end{align}
with $\textbf{X}_{\OUT}[\omega] = \textbf{s}[\omega] \textbf{X}_{\IN}[\omega]$ and the phase
$\phi = 2 \arg[\kappa +  2 i \omega ]$. The frequency dependent gain $\mathcal G [\omega]$
contains the zero-frequency amplitude gain $\sqrt{\mathcal G_{Q}} = 8  G / \kappa$.
Here the P-quadratures contain the amplified X-quadratures of the respective other mode,
where for gain we require $ G > \kappa/8 $. 
Note, although the scattering matrix in this basis seems somewhat asymmetric, we
still have a reciprocal amplifier. Transforming back to the non-hermitian
basis of $d_n$ operators we obtain a completely symmetric scattering matrix.
The QND amplifier has as well no gain-bandwidth limit, but the bandwidth of
$ D_{QND} = \sqrt{\sqrt{2}-1} \kappa$ is half as large as for the OIBA.

To determine the added noise for the QND-amplifier we calculate
the symmetrized noise spectral density and obtain for
the added noise  (referred back to the input)
\begin{align}
 \bar n_{\ADD} = \frac{1}{ \mathcal G_{Q}  } \left( \frac{1}{2} + \bar n^{T} \right),
\end{align}
here $\bar n^{T}$ denotes the thermal bath which the cavity receiving the input signal
is coupled to. Clearly we can reach the quantum limit for large gain, even at finite temperature.

Alternatively we can start out from the detuned single-mode DPA
\begin{align}
 \hH = \Delta \hat a^{\dag} \hat a
  + \frac{\nu}{2}
  \left( \hat a^{\dag} \hat a^{\dag} e^{ -i \varphi }
      + \hat a \hat a  e^{ + i \varphi }\right),
\end{align}
where we accounted as well for a possible phase $\varphi$ determining which quadratures is squeezed and which is amplified. We now choose balanced detuning and coupling strength, i.e.,  $\Delta = \nu$, and obtain the for the output of an aribitary quadrature in frequency space 
\begin{align}
    \hat X^{\theta}_{\OUT}[\omega]
 =&  
   - e^{i 2 \phi[\omega] }
   \hat X^{\theta}_{\IN}[\omega]
   -
    \frac{ 2 \kappa   \nu
          \sin\left[  \frac{\theta - \varphi}{2}\right]}
 {\left[- i \omega + \frac{\kappa}{2}  \right]^2  }
     \hat X^{\varphi}_{\IN}[\omega],
\end{align}
with $\hat X^{\theta } = (\hat a e^{i\frac{\theta}{2}} +\hat a^{\dag} e^{-i\frac{\theta}{2}}  )/\sqrt{2}$
and $\phi[\omega] = \arg{[i\omega + \frac{\kappa}{2}}]$. An input quadrature with phase $\theta = \varphi$ is simply reflected, while the corresponding orthogonal quadrature with phase $\theta = \varphi \pm \pi$ contains the amplified input quadrature $\hat X^{\varphi }_{\IN}$. The corresponding amplitude gain scales linearly with $\nu$ and is independent of the bandwidth, which is fixed at $\kappa \sqrt{ \sqrt{2} - 1  }$. The added noise on resonance is not zero, but scales inversly with the gain. However, a drawback is that the output of the amplified quadrature contains as well an unddesired mixing of both quadratures which is due to the finite reflection with phase $\phi[\omega]$.

\section{OIBA: Influence of asymmetries}\label{AppAsymmetries}
%
\begin{figure} 
  \centering\includegraphics[width=\columnwidth]{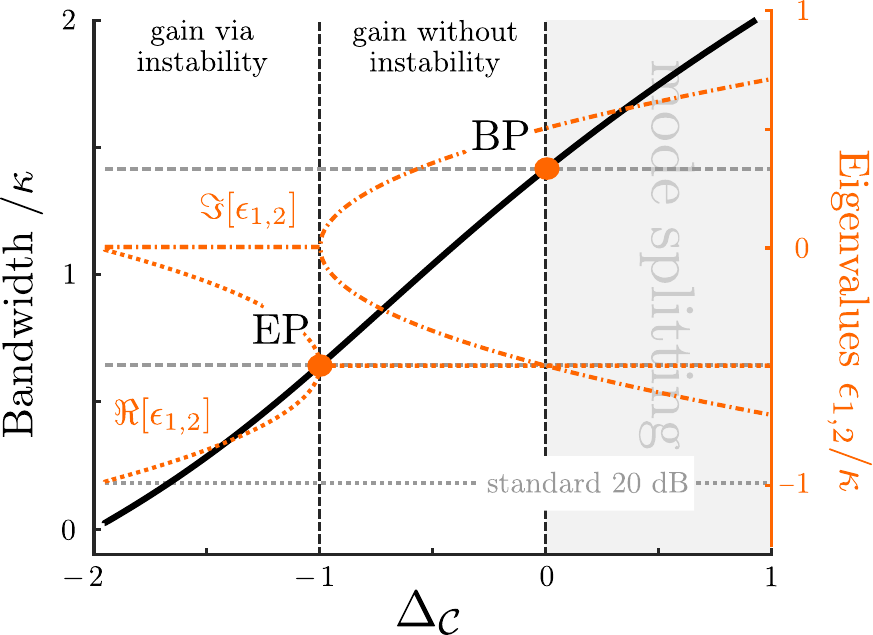} 
 	\caption{Bandwidth (black solid line) and real and imaginary part of the eigenvalues (orange dashed-dotted and dotted line) of the OIBA as a function of relative cooperativities $\Delta_{\mathcal C} = \mathcal C_{2} - \mathcal C_{1} - 1$ for symmetric decay rates. For a balanced choice of coupling parameters, i.e.,  the QND case for $\mathcal C_{2} = \mathcal C_{1}$, the eigenvalues are degenerate and the system has an exceptional point (EP). For $\Delta_{\mathcal C} = 0$ the system reaches the Bogoliubov point (BP) of maximal bandwidth, beyond this point the gain profile splits into two peaks. This behavior is independent of the gain value, for comparison the dashed dotted line shows the bandwidth of a standard DPA at 20dB. The OIBA clearly outperforms the latter case. } 
 	\label{Figure9}
\end{figure}
In this section we briefly discuss the influence of asymmetries in the
decay rates of the two cavity modes ($\kappa_{1} \neq \kappa_2$) for the OIBA, as well as deviations from the optimal matching condition. Starting with the eigenvalues of the dynamical matrix (involving coherent and dissipative evolution) of the system   
\begin{align}
 \epsilon_{1,2} =& - \frac{\kappa_{+}}{2} \pm  i \sqrt{ \widetilde G^2 - \frac{\kappa_{-}^2}{4} },
 \hspace{0.2cm}
 \kappa_{\pm} = \frac{\kappa_{1} \pm \kappa_{2}}{2},
\end{align}
with $\widetilde G = \sqrt{G_{2}^2 - G_{1}^2}$ as used in the main text. The eigenvalues for symmetric decay rates are depicted in Fig.\ref{Figure9} together with the bandwidth of the amplifier, showing that the point of optimal bandwidth appears when the absolute value of the real and imaginary part of the eigenvalues coincide. We call this point the Bogoliubov point. 
The latter point of optimally imbalanced parametric processes persists also for asymmetric decay rates, we can extract the condition from the eigenvalues as
\begin{align}
\widetilde G =  \frac{\sqrt{\kappa_{+}^2 +  \kappa_{-}^2}}{2}
\hspace{0.2cm}  \Rightarrow \hspace{0.2cm}
\Delta_{\mathcal C}  =  
\frac{ \kappa_{-}^2  }{2\kappa_1 \kappa_2} 
\ ,
\end{align} 
with the definition $\Delta_{\mathcal C}      =   \left[   \mathcal C_{2} - \mathcal C_{1} - 1 \right]$
and  the cooperativities $\mathcal C_n = 4 G_n^2/ (\kappa_1 \kappa_2)$. For the case of symmetric decay rates $\kappa_{-} = 0$  we can characterize the  deviations from the matching condition for the Bogoliubov amplifier as $\Delta_{\mathcal C} \neq 0$.
However, an asymmetry in the decay rates does only modify the condition for the Bogoliubov point and the scattering behaviour gets slightly modified,
the scattering matrix in the quadrature basis becomes
\begin{align}
 \textbf{s}[\omega] =&  
\left(
\begin{array}{cccc}
\mathcal R_{-}[\omega]   & 0 &  0    & \mathcal T_{-}[\omega] 
 \\
 0 &\mathcal R_{-}[\omega]  &  \mathcal T_{+}[\omega]  &  0  
 \\
  0  &   \mathcal T_{-}[\omega] & \mathcal R_{+}[\omega]  & 0 
 \\
  \mathcal T_{+}[\omega] &   0 & 0 & \mathcal R_{+}[\omega] 
 \\
\end{array}
\right),
\end{align}
with the reflections and transmissions
\begin{align}
 \mathcal R_{\pm}[\omega] =& - \frac{\mathcal C_{1} - \mathcal C_{2}  +   \left[ 1 \mp i \frac{ 2 \omega}{\kappa_1}    \right] \left[ 1  \pm  i \frac{ 2\omega}{\kappa_2}     \right]}
                         {   \mathcal C_{2} - \mathcal C_{1} +    \left[ 1 - i \frac{ 2 \omega}{\kappa_1}    \right]\left[ 1 - i \frac{ 2 \omega}{\kappa_2}   \right]},
     \nonumber \\
 \mathcal T_{\pm}[\omega] =& + \frac{ 2 \left[ \sqrt{\mathcal C_{1}}  \pm \sqrt{\mathcal C_{2}}  \right]   }
                         { \mathcal C_{2} - \mathcal C_{1} +    \left[ 1 - i \frac{ 2 \omega}{\kappa_1}    \right] \left[ 1 - i \frac{ 2 \omega}{\kappa_2}   \right]},
\end{align}
with $\mathcal T_{+}[\omega]$ denoting the phase sensitive gain as a function of frequency, while $\mathcal T_{-}[\omega]$ describes the corresponding deamplification of the orthogonal quadrature.
On resonance the transmission and reflection  become 
\begin{align}
 \mathcal R_{\pm}[0] =   \frac{\Delta_{\mathcal C}  }
                         { \Delta_{\mathcal C}   + 2},
     \hspace{0.3cm}
 \mathcal T_{\pm}[0] =  + \frac{ 2 \left[ \sqrt{\mathcal C_{1}}  \pm \sqrt{\mathcal C_{2}}  \right]   }
                         {\Delta_{\mathcal C}  + 2},    
\end{align}
thus, the reflection always vanishes for $\Delta_{\mathcal C} = 0$, which is possible achive for  the impedance matching condition
$\widetilde G = \sqrt{\kappa_{1} \kappa_{2}}/2$. But this does not coincide with the point of optimal imbalance for finite $\kappa_{-}$, i.e., $\Delta_{\mathcal C}  = 2\kappa_{-}^2/ (\kappa_{1} \kappa_{2})$. Meaning that the condition for the optimal bandwidth and 
and the impedance matching condition only coincide for symmetric decay rates. 
Moreover, the power gain $|\mathcal T_{+}[0]|^2$ increases with the cooperativities as long as $\Delta_{\mathcal C} \geq -1 $, after the exceptional point, see Fig.\ref{Figure9}. For $ -2 < \Delta_{\mathcal C} < -1$ the denominator is smaller than one and thus the resulting gain is enhanced and the reflected signal starts to get become amplified as well.

\section{Nonreciprocal Bogoliubov amplifier}\label{AppNonreciprocal}
The Bogoliubov amplifier introduced in this work overcome multiple shortcoming in comparison to the standard operation mode of parametric amplifiers. However, one drawback remains, as the introduced (two-mode) Bogoliubov amplifiers are all reciprocal, i.e., they amplify signals in both directions. Consequently, in a real architecture they would have to be combined with nonreciprocal elements such as circulators to protect the signal source from unwanted (amplified) noise coming from higher up in the measurement chain. Circulators are noisy and bulky elements, hence to remove their necessity we discuss in this section how we can promote a reciprocal two-mode Bogoliubov amplifier to a nonreciprocal amplifier. 

For this we apply the recipe introduced in our former work \cite{Metelmann2015} to the OIBA. The method is based on balancing the coherent interaction with the corresponding dissipative interaction. Such a dissipative interaction can be engineered via coupling
the two main modes in the OIBA indirectly via a damped auxiliary mode. 
In the Bogoliubov basis we stay in the realm of stable dynamics and simply include hopping among all three modes. The Hamiltonian
of the full system yields  
\begin{align}\label{EqHamCircBog}
 \hH =    \lambda \left(  \hat \beta_{1}^{\dag} \hat \beta_{2}  
        +   \hat \beta_{3}^{\dag} \hat \beta_{1}  
        +   \hat \beta_{3}^{\dag} \hat \beta_{2} e^{i\varphi}  
        + h.c. \right),
\end{align}
where we assumed uniform hopping strengths $\lambda$ among the Bogoliubov modes for simplicity, and mode-3 is our auxiliary mode. Coupling these modes to input/output ports with strength $\kappa$, and setting $\lambda = \frac{\kappa}{2}$, as well as $\varphi = -\frac{\pi}{2}$, we obtain the scattering  behaviour of a circulator on resonance:
\begin{align}
\left(
\begin{array}{c}
 \hat \beta_{1,\OUT}[0]
 \\
 \hat  \beta_{2,\OUT}[0] 
 \\
 \hat  \beta_{3,\OUT}[0] 
\end{array}
\right)
=
 \left(
\begin{array}{ccc}
 0 & 0 & i \\
 i & 0 & 0 \\
 0 & 1 & 0 
\end{array}
\right)
\left(
\begin{array}{c}
 \hat \beta_{1,\IN}[0]
 \\
 \hat  \beta_{2,\IN} [0]
 \\
 \hat  \beta_{3,\IN}[0] 
\end{array}
\right),
\end{align}
realizing clockwise circulation ($1\rightarrow2\rightarrow3$) of signals among the three ports, while changing the phase by $\pi$  leads to circulation in the opposite direction. Next we perform a transformation from the basis of the local Bogoliubov modes $\hat \beta_n$ into the basis of quadrature operators $\hat X_{n}/\hat P_n$ of the three modes, i.e. we insert
\begin{align}
 \hat \beta_{n} =& \cosh r \hat a_{n} + \sinh r \hat a_{n}^{\dag}
 \nonumber \\
 =&
 \frac{1}{\sqrt{2}} 
 \left[ \cosh r  
   \left( \hat X_{n} + i \hat P_{n}\right) 
   +  \sinh r  
   \left( \hat X_{n} - i \hat P_{n}\right)\right]
   \nonumber \\ 
   =&
 \frac{1}{\sqrt{2}} 
 \left[   \hat X_{n} e^{r} 
     + i \hat P_{n} e^{-r}  \right]
   ,
 \hspace{0.5cm}
 n \in 1,2,3,
\end{align}
into the Hamiltonian in Eq.(\ref{EqHamCircBog}). For $\phi = \pm \frac{\pi}{2}$ we obtain
\begin{align}
 \hH
      =&  \; \; \; \;
      \lambda
      \left[
      \hat X_{1}\hat X_{2} e^{2r}
      + \hat P_{1} \hat P_{2} e^{-2r}
                      \right]
                      \nonumber \\ &
      +   \lambda
      \left[  \hat X_{3}
       \left(\hat X_{1} e^{2r} \mp \hat P_{2} \right)
      + \hat P_{3} \left(
         \hat P_{1}    e^{-2r} \pm \hat X_{2} \right)
                      \right],
\end{align}
such an interaction requires five parametric processes among the three modes, i.e., two-mode squeezing and frequency conversion between mode pairs $(1-2)$ and  $(1-3)$, and in addition one hopping process between mode-2 and mode-3. From the Hamiltonian in the quadrature basis we can already see that an input in the X-quadratures will be amplified, while the P-quadrature will be squeezed. Including the input/output ports the quantum Langevin equations for the system become
\begin{align}
    \frac{d}{dt} \hat X_{1} 
    =& -\frac{\kappa + \Gamma}{2} \hat X_{1}
       + \left[ \lambda  \pm \frac{\Gamma}{2}\right]
       e^{-2r} \hat P_{2}
       \nonumber \\ &      
       -  \sqrt{\Gamma}
       e^{-2r}  \hat P_{3,\IN} 
       - \sqrt{\kappa} \hat X_{1,\IN},
       \nonumber \\
     \frac{d}{dt} \hat P_{1} 
    =& -\frac{\kappa + \Gamma}{2} \hat P_{1}
       - \left[ \lambda  \pm \frac{\Gamma}{2}\right]
        e^{+2r} \hat X_{2} 
       \nonumber \\&  
       + \sqrt{\Gamma} 
        e^{+2r} \hat X_{3,\IN} 
       - \sqrt{\kappa} \hat P_{1,\IN},  
       \nonumber \\
    \frac{d}{dt} \hat X_{2} 
    =& -\frac{\kappa + \Gamma}{2} \hat X_{2}
       + \left[ \lambda \mp \frac{\Gamma}{2} \right]
       e^{-2r} \hat P_{1} 
       \nonumber \\ & 
        \pm \sqrt{\Gamma} \hat X_{3,\IN}
       - \sqrt{\kappa} \hat X_{2,\IN},
       \nonumber \\
     \frac{d}{dt} \hat P_{2} 
    =& -\frac{\kappa + \Gamma}{2} \hat P_{2}
       - \left[ \lambda \mp \frac{\Gamma}{2} \right]
        e^{+2r} \hat X_{1} 
       \nonumber \\ & 
         \pm \sqrt{\Gamma} \hat P_{3,\IN} 
       - \sqrt{\kappa} \hat P_{2,\IN},     
\end{align}
with $\Gamma = 4\lambda^2/\kappa_3$. Here we already adiabatically eliminated the auxiliary mode assuming it is overdamnped with rate $\kappa_3$. The latter is not a necessary step for the nonreciprocal amplifier to work on resonance, however, it allows for a reduced description here. Moreover, assuming an overdamped auxiliary mode enhances the bandwidth over which the amplifier will be unidirectional \cite{Kamal2017}.  The phase $\varphi$ decides in which direction the amplifier will work, taking for example $\varphi = +\frac{\pi}{2}$ we obtain the following scattering matrix on resonance
\begin{align}
\left(
\begin{array}{c}
 \hat X_{1,\OUT}[0]
 \\
 \hat  P_{1,\OUT}[0] 
 \\
 \hat  X_{2,\OUT}[0] 
  \\
 \hat  P_{2,\OUT}[0] 
\end{array}
\right)
=
 \left(
\begin{array}{cccc}
 0 & 0 & 0 & 0 \\
 0 & 0 & 0 & 0  \\
 0 & e^{-2r} & 0 & 0  \\
e^{+2r}& 0 & 0 & 0 
\end{array}
\right)
\left(
\begin{array}{c}
 \hat X_{1,\IN}[0]
 \\
 \hat P_{1,\IN} [0]
 \\
 \hat X_{2,\IN}[0] 
  \\
 \hat P_{2,\IN}[0] 
\end{array}
\right),
\end{align}
where the an input in mode 1 will show up amplified and squeezed in the output of mode 2, while no input on mode 2 will show up at the output of mode 1. Hence we have a nonreciprocal amplifier with no gain-bandwidth limit and no back-reflections. 
The power gain and added noise of the amplifier as a function of frequency become
\begin{align}
\mathcal G[\omega] =   \frac{e^{+ 4r} }{\left[1 +  \frac{\omega^2}{\kappa^2} \right]^2},
\hspace{0.2cm}
     \bar n_{\ADD}^{-}[\omega]
     = e^{-4r}  \frac{\omega^2}{\kappa^2}\left[2 +  \frac{\omega^2}{\kappa^2} \right]  ,
 \end{align}
 reaching the quantum limit of zero added noise not only on resonance (i.e. for $\omega = 0$), but approaching this limit as well for finite frequency. The reason for this is that the fluctuations originating from the auxiliary cavity do not show up amplified at the output port of mode 2. However, the drawback is, that the noise of the mode 3 will show up amplified at the output of mode 1. The symmetrized output noise spectra of mode 1 become
\begin{align}
 2 \bar S_{X_1 X_1/P_1 P_1}^{-}[\omega]
 =&
  \frac{  \frac{\omega^2}{\kappa^2}  +  e^{\mp 4r} }
        {\left[1 + \frac{\omega^2}{\kappa^2} \right]},
\end{align}
which does not vanish even on resonance. To circumvent this we can operate the amplifier in the opposite direction, i.e., for $\varphi = + \frac{\pi}{2}$ signals entering mode 2 will show up amplified in mode 1. For this direction the properties of the gain and the bandwidth stay the same, while the amplified noise of the auxiliary mode will then leak as well out of port 1. This will modify the added noise of the amplification process as well as the noise contributions routed to port 2, i.e. we obtain 
\begin{align}
 2 \bar S_{X_2 X_2/P_2 P_2}^{+}[\omega]
 =& 1,
\hspace{0.2cm}
\bar n_{\ADD}^{+}[\omega] = 
       \left[\frac{\omega^2}{\kappa^2} + \frac{\omega^4}{\kappa^4}  \right] 
         e^{- 4r} 
         + \frac{\omega^2}{\kappa^2} ,
\end{align}
meaning that no amplified noise will show up on the port the signal is injected too, and the drwaback for the added noise is minor, i.e., the amplifier is still quantum limited on resonance but is finite away from it. Here we see that the routing of fluctuations in a nonreciprocal system can be a powerful tool, which can be exploited to enhance the noise properties of the amplifier, but can as well be used to generate pure entanglement in the presence of thermal fluctuations. \cite{Orr2023}.
 

\end{document}